\documentclass[aps,twocolumn,floatfix, showkeys, superscriptaddress, nofootinbib]{revtex4-2}
\usepackage{amsmath}
\usepackage{graphicx,bm,hhline, xcolor, float, xcolor}
\usepackage[colorlinks]{hyperref}

\newcommand{\br}{{\bm r}}

\begin{document}
\title{Dense Plasma Opacity from Excited States Method}

\author{C. E. Starrett}
\email{starrett@lanl.gov}
\affiliation{Los Alamos National Laboratory, P.O. Box 1663, Los Alamos, NM 87545, U.S.A.}

\author{C. J. Fontes}
\affiliation{Los Alamos National Laboratory, P.O. Box 1663, Los Alamos, NM 87545, U.S.A.}

\author{H. B. Tran Tan}
\affiliation{Los Alamos National Laboratory, P.O. Box 1663, Los Alamos, NM 87545, U.S.A.}

\author{J. M. Kasper}
\affiliation{Los Alamos National Laboratory, P.O. Box 1663, Los Alamos, NM 87545, U.S.A.}

\author{J. R. White}
\affiliation{Los Alamos National Laboratory, P.O. Box 1663, Los Alamos, NM 87545, U.S.A.}
\affiliation{Department of Astronomy, University of Texas at Austin, Austin, TX 78712, U.S.A.}

\date{\today}
\begin{abstract}
The self-consistent inclusion of plasma effects in opacity calculations is a significant modeling challenge.  As density increases, such effects can no longer be treated perturbatively.  Building on a a recently published model that addresses this challenge, we calculate opacities of oxygen at solar interior conditions.  The new model includes the effects of treating the free electrons consistently with the bound electrons, and the influence of free electron energy and entropy variations are explored.  It is found that, relative to a state-of-the-art-model that does not include these effects, the bound free-opacity of the oxygen plasmas considered can increase by 10\%.
\end{abstract}
\maketitle

\section{Introduction}
The opacity of dense plasmas is an important quantity with applications in many fields, including solar astrophysics \cite{basu2008helioseismology}, white dwarfs modelling~\cite{saumon2022current}, and inertial confinement fusion (ICF) \cite{atzeni04}.  However, accurate calculation of opacity remains challenging since it requires the merging of atomic and plasma physics, the inclusion of multiple, significant physical phenomena such as continuum lowering \cite{ciricosta2012direct, vinko12}, line shapes \cite{gomez2022introduction}, multiple scattering \cite{shaffer22dense}, and self-consistent plasma effects \cite{hu2022probing}, as well as the proper accounting for a multitude of excited states \cite{colgan16, starrett24excited, iglesias1996updated,bar1989super}.

Many opacity models start with isolated atoms or ions and obtain the ion species and excited state populations by solving the set of collisional-radiative equations or assuming Local Thermodynamic Equilibrium (LTE)~\cite{colgan16, iglesias1996updated}. Plasma effects on electronic structure are then accounted for in an \textit{ad hoc} manner. Some methods, such as the superconfiguration approach \cite{bar1989super, gill2023superconfiguration, blenski2000superconfiguration, pain17}, directly include plasma effects in the Hamiltonian \cite{faussurier18, hansen2023self, hansen2006comparison, ovechkin2014reseos, pain2006self, pain2003self}.  In some of these methods, a Thomas-Fermi treatment of the continuum is used, with the resulting discontinuities caused by pressure ionization being smoothed by the statistical treatment of (super)configurations.  A formal theory for superconfigurations was introduced in \cite{blenski07}; however, it has not been possible, to date, to implement this method.  

In Ref.~\cite{starrett24excited}, a variational method that consistently treats bound- and free-electron excited states was introduced.  This method was used in Ref.~\cite{thelen24} to predict the energies of excited states, and the results were in good agreement with experiments~\cite{ciricosta2012direct}. In this paper, we demonstrate the capability of the variational method~\cite{starrett24excited} in predicting plasma opacities. Furthermore, whereas the atomic model used in Ref.~\cite{starrett24excited} assigns spheres of a fixed radius for different atoms or ions with different excited states, in this work we remove this restriction by allowing the sphere size to vary with excited state.



The outline of this paper is as follows.  In Sec.~\ref{sec_esm} we review the excited states method (ESM) \cite{starrett24excited}.  A summary of the atomic model in ESM is given in Sec.~\ref{sec_os}, which restricts all excited states to have the same atomic sphere size.  A key concept in the ESM is choosing the one-electron occupation factors.  In Sec.~\ref{sec_nxi} we give a simple example of this concept.  In Sec.~\ref{sec_vs} a variation on the atomic model in which the atomic sphere size varies with excited state is introduced.  In Sec.~\ref{sec_calca} we discuss the method for calculating the absorption coefficient (opacity) from the ESM. The results of our calculations are presented and discussed in Secs.~\ref{sec_oxy} and~\ref{sec_si}. Finally, conclusions are drawn in Sec.~\ref{sec_con}.

\section{Excited States Method\label{sec_esm}}
The excited states method (ESM), as described in Ref.~\cite{starrett24excited}, is a variational method capable of determining the excited state energies, wavefunctions, and populations. In the ESM, an excited state $x$ is approximated by a system with one-electron orbitals $\psi_{xi}$ and a set of numbers $n_{xi}$ representing the occupation factors of these orbitals. Among the orbitals included within an excited state, one differentiates between \textit{core} orbitals with integer occupation numbers (this number can be zero) and \textit{hybridized} orbitals with fractional occupation factors.  A statistical ensemble of such excited states is constructed in this way and the total ensemble free energy obtained. The minimization of this free energy, subject to certain constraints, allows one to find the one-electron energies $\epsilon_{xi}$, one-electron orbitals $\psi_{xi}$, and the probability $W_x$ of excited states $x$. 

As a starting point for our discussion on using ESM to compute opacities, some details from Ref.~\cite{starrett24excited} are reproduced here. We will only present the equations obtained from the variational procedure; the formulation of the free energy (as well as the constraints) may be found in Ref.~\cite{starrett24excited}. Moreover, we adopt the atomic model of reference \cite{starrett24excited}, which introduces a simplification by considering the plasma as a collection of independent charge-neutral spheres, each with a nucleus of charge $Z_x$ and $Z_x$ electrons representing an atom in an excited state $x$.  As in Ref.~\cite{starrett24excited}, all spheres have the same volume, taken as the average volume per atom in the plasma.  This, in turn, is set by input of mass density and atomic weight\footnote{We shall discuss how to relax this condition in Sec.~\ref{sec_vs}.}. Spherical symmetry is assumed within each sphere.

\subsection{The ESM atomic model}\label{sec_os}
The minimization of the free energy with respect to the one-electron orbitals 
\begin{equation}
    \psi_{x\epsilon l}(\br) = y_{x\epsilon l}(r)Y_{lm}(\hat{\bm r})/r
\end{equation}
gives the one-particle Schr\"odinger equation
\begin{align}\label{eq:Schrodinger_eqn}
    &\left[-\frac{1}{2}\frac{\partial^2}{\partial r^2}+ \frac{l(l+1)}{2r^2}+V^{\rm el}_x(r)+V^{\rm xc}_x(r)-\gamma\right]y_{x\epsilon l}(r)\nonumber\\
    &=\epsilon_{xl} y_{x\epsilon l}(r)\,,
\end{align}
where $l$ is the orbital angular momentum. Here, $V^{\rm el}_x(r)$ is the electrostatic potential given by
\begin{align}
    V_x^{\rm el}(r) =-\frac{Z_x}{r}  + \int d\br' \frac{n_x(r')}{|\br -\br'|}\,,
\end{align}
where
\begin{equation}\label{eq:def_nx}
    n_x(r)=\sum_l2(2l+1)\int d\epsilon|\psi_{x\epsilon l}(r)|^2\,,
\end{equation}
is the electron density\footnote{The integral over $\epsilon$ in Eq.~\eqref{eq:def_nx} may be understood as containing both a discrete summation over bound states and an integration over the continuum.}. Note that $V_x^{\rm el}(r)$ may be derived from an electrostatic energy, $V^{\rm el}_x(r)=\delta E^{\rm el}_x/\delta n_x(r)$, where
\begin{align}\label{eq:Eel}
    E_x^{el} = \frac{1}{2} \int d{\bm r}d{\bm r}'\frac{n_x(r) n_x(r')}{|\br -\br'|}-  Z_x \int d{\bm r} \frac{n_x(r)}{r}\,.
\end{align}
Similarly, the exchange and correlation potential $V^{\rm xc}_x(r)$ may be derived from an exchange and correlation energy $E^{\rm xc}_x$ via $V^{\rm xc}_x(r)=\delta E^{\rm xc}_x/\delta n_x(r)$. In this work, we assume the local density approximation (LDA) wherein $E^{\rm xc}_x$ is a functional of $n_x(r)$ alone
\begin{equation}\label{eq:Exc}
    E^{\rm xc}_x=\int d\br \epsilon^{\rm xc}[n_x(r)]\,.
\end{equation}
Finally, $\gamma$ is a constant setting the origin of energy which arises from demanding charge neutrality for the whole ensemble (see Eq.~(8c) in Ref.~\cite{starrett24excited}).

The minimization of the free energy with respect to the occupation number $n_{xi}$ gives
\begin{equation}\label{eq:n_FD}
    n_{xi}=\frac{1}{{\exp\left[(\epsilon_{xl}-\mu_{x\epsilon l})/T\right]+1}}\,,
\end{equation}
where $T$ is the plasma temperature and $\mu_{x\epsilon l}$ is determined by requiring that $n_{xi}$ matches some input $f_{xi}$. The quantities $f_{xi}$ may be chosen arbitrarily as long as they satisfy $0\leq f_{xi}\leq 1$ (see Eq.~(8d) in Ref.~\cite{starrett24excited}).

Finally, minimizing the ensemble free energy with respect to the excited state probability $W_x$ gives
\begin{equation}\label{eq:W_x}
    W_x=\frac{\exp(-F_x/T)}{\sum_x\exp(-F_x/T)}\,,
\end{equation}
where $F_x=E_x+TS_x$ is the internal free energy of the excited state $x$. Here, $E_x$ is the excited state internal energy given by
\begin{equation}
    E_x = E_x^{(0)} + E_x^{\rm el} + E_x^{\rm xc}\,,
\end{equation}
where 
\begin{align}\label{eq:kinetic_energy}
E_x^{(0)} &=\sum_l2(2l+1)\int d\epsilon\,n_{x\epsilon l}\nonumber\\
&\times\int dry_{x\epsilon l}^*(r)\left[-\frac12\frac{\partial^2}{\partial r^2}+\frac{l(l+1)}{2r^2}\right] y_{x\epsilon l}(r)\nonumber\\
&=\sum_l2(2l+1)\int d\epsilon\,n_{x\epsilon l}\nonumber\\
&\times\int dry_{x\epsilon l}^*(r)\left[\epsilon_{x\epsilon l}+ \gamma - V_x^{\rm el}(r)- V_x^{\rm xc}(r)\right] y_{x\epsilon l}(r)\,,
\end{align}
is the kinetic energy, whereas the electrostatic and exchange-correlation energies $E_x^{\rm el}$ and $E_x^{\rm xc}$ are given in Eqs.~\eqref{eq:Eel} and~\eqref{eq:Exc}, respectively.

Finally, the internal entropy $S_x$ of the excited state $x$ may be split into three parts
\begin{equation}
    S_x = S_x^{\rm c} + S_x^{\rm e} + S_x^{\rm xc}\,,
\end{equation}
where $S_x^{\rm c}$ represents the entropy of one-electron core (c) orbitals with integer occupation numbers, $S_x^{\rm e}$ comes from excited (e) one-electron orbitals with fractional occupation, and finally $S_x^{\rm xc}$ is due to exchange and correlation effects. More explicitly, we have
\begin{equation}
    S_x^c = \ln g_x \label{eq_sc}\,,
\end{equation}
where $g_x$ is the total number of core microstates in $x$, given by
\begin{equation}\label{eq:gx}
    g_x = \prod_{\epsilon,l} \frac{d_{x\epsilon l} !}{(d_{x\epsilon l} n_{x\epsilon l}) ! (d_{x\epsilon l} - d_{x\epsilon l}n_{x\epsilon l})!}\,,
\end{equation}
with $d_{x\epsilon l}=2(2l+1)$ being the degeneracy of the orbital $\psi_{x\epsilon l}$ and, as indicated, the product runs over all one-electron core orbitals contained in $x$.  The fractional-occupation contribution $S_x^{\rm e}$ is given by the mean-field entropy expression
\begin{align}\label{eq:S_x^fd}
    S_x^{\rm e}&=-\int dr\sum_{l}2(2l+1)\int d\epsilon|y_{x\epsilon l}(r)|^2\nonumber\\
    &\times\left[n_{x\epsilon l}{\,\rm ln\,}n_{x\epsilon l}+(1-n_{x\epsilon l}){\rm ln}(1-n_{x\epsilon l})\right]\,,
\end{align}
where the integral over $\epsilon$ should be understood as containing both a discrete sum over bound states and an integration over the continuum, but restricted only to the fractionally occupied orbitals in $x$. Note that the entropy treatment in this work is treated sightly different from that presented in Ref.~\cite{starrett24excited}. Here we do not use the mean field entropy expression for $S_x^{\rm c}$, but rather use Eqs.~\eqref{eq_sc} and~\eqref{eq:gx}.  The mean field expression~\eqref{eq:S_x^fd} arises from Eq.~\eqref{eq_sc} in the limit of large $d_{x\epsilon l}n_{x\epsilon l}$ by using Sterling's approximation for $\ln n!$.  This is an excellent approximation for large $n$, but is poor for small $n$, as is the case, for example, for low-lying shells.  

We have now obtained all the ingredients of our ESM. The procedure for a practical computation is as follows. First, one chooses a set of occupation factors $f_{xi}$ (integer and fractional) that defines the excited states.  One then solves the one-particle Schr\"odinger equation~\eqref{eq:Schrodinger_eqn} for the orbitals and eigenvalues associated in a self-consistent manner. The different excited states are connected though the energy offset $\gamma$ , which can be determined iteratively. Once the excited state energies $E_x$ and entropies $S_x$ are determined, the probabilities $W_x$ are found with Eq.~\eqref{eq:W_x}. It should be clear that a physically sensible choice for the occupation numbers $f_{xi}$ is critical for an accurate ESM calculation. A systematic method to choose these number is described in Ref.~\cite{starrett24excited} and shall not be presented here. Instead, we consider in Sec.~\ref{sec_nxi} a simple example to demonstrate this method.

\subsection{A simple example}\label{sec_nxi}
Let us now demonstrate the ESM with a simple example. For definiteness, we consider an oxygen plasma at conditions near the boundary of the solar radiative/convective zones, i.e., a mass density of 0.11 g/cm$^3$ and a temperature of 175 eV ($\approx$ 2 MK). We concentrate on three explicit excited states with integer occupations of the $1s$ shell
\begin{enumerate}
    \item 1s$^0$ + FD\,,
    \item 1s$^1$ + FD\,,
    \item 1s$^2$ + FD\,.
\end{enumerate}
This notation means, for example, that in configuration 1, there are no electrons present on the $1s$ shell, while the remaining 8 electrons (for oxygen) are in higher energy states and have Fermi-Dirac (FD) occupation, as in Eq.~\eqref{eq:n_FD}. Note that the Kohn-Sham, density functional theory based average atom model \texttt{Tartarus}~\cite{starrett19} indicates that there are bound states up to and including $n=4$, so a more complete model would have integer occupations up to the $n=4$ shells.  For the sake of simplicity, here we limit ourselves only to the $n=1$ shell.
\begin{table}
\begin{center}
\bgroup
\def\arraystretch{1.5}%
\begin{ruledtabular}
\begin{tabular}{cccc}
 &One Sphere& Variable Spheres & \texttt{Tartarus}\\
$W_x$ : 1s$^0$+FD & 0.6937 & 0.6841 & \\
$W_x$ : 1s$^1$+FD & 0.2914 & 0.2981 & \\
$W_x$ : 1s$^2$+FD & 0.0149 & 0.0179 & \\
\vspace{0.3cm}
Ave. 1s        & 0.3212 & 0.3339 & 0.3707 \\
$n_e^0$ [$a_0^{-3}$]       & 4.52$\times 10^{-3}$ & 4.52$\times 10^{-3}$ & 4.50$\times 10^{-3}$ \\
$P_e$ [Mbar]               & 8.55 & 8.55 & 8.50 \\
\end{tabular}
\end{ruledtabular}
\egroup
\end{center}
\caption{Excited states probabilities $W_x$, average electron density $n_e^0$, and electron pressure $P_e$ for an oxygen plasma at 0.11 g/cm$^3$ and 175 eV as computed with ESM with a ``One-sphere'' model (see Sec.~\ref{sec_os}), with ESM with a ``Variable-spheres'' model (see Sec.~\ref{sec_vs}), and with the DFT-based average atom model \texttt{Tartarus}~\cite{starrett19}. Only three configurations with core electrons on the $1s$ shell are considered. The label ``Ave. $1s$'' refers to the average occupation of the $1s$ orbital.\label{tab_smallex}}
\end{table}

In Table~\ref{tab_smallex}, we present the probabilities $W_x$, electron density $n^0_e$, and electron pressure $P_e$ as predicted by our ESM summarized in Sec.~\ref{sec_os}, which we are calling the One-sphere model (as opposed to the Variable-spheres, discussed later in Sec.~\ref{sec_vs}). Presented also is the average $1s$ occupation, electron density, and electron pressure as predicted by the DFT-based average atom model \texttt{Tartarus}. Note that in contrast to ESM, \texttt{Tartarus} cannot predict the probabilities of the individual configurations. One observes that the DFT result for the average occupation of the 1s shell is $\sim$ 10 \% higher than that from the One-Sphere model, but the two approaches agree on $n_e^0$ and $P_e$ to within 1 \%.  This behavior reflects the fact that, although the DFT model is designed to get the correct electron density and average energy, its predictions for individual orbitals are not strictly physically meaningful.  In contrast, with the ESM, one aims at getting the individual excitation energies, as well as the electron density and pressure, correct.
\begin{table}
\begin{center}
\bgroup
\def\arraystretch{1.5}%
\begin{ruledtabular}
\begin{tabular}{ccc}
1s$^0$+FD & 1s$^1$+FD & 1s$^2$+FD  \\
0.1038 & 0.1183 & 0.1375 \\
\end{tabular}
\end{ruledtabular}
\egroup
\end{center}
\caption{Effective mass densities (g/cm$^3$) of the excited states for a plasma mass density of 0.11 g/cm$^3$.\label{tab_vs}}
\end{table}

For reference, in Table \ref{tab_vs} we show the effective mass densities of the three excited states.  This number corresponds to the mass density that the plasma would have if it were only composed of the excited state atom at the determined volume.

We have presented an example for which our ESM is applied. Although the example shows that the original ESM works well, improvements to the model are possible. One such improvement can be made by realizing that, physically, ions of different charges will repel their neighboring ions differently. This effect is not reflected in the original ESM, which assumes that all excited state spheres are of the same volume. Since an ion with higher charge will, on average, have neighboring atoms farther away from its center than one with a lower charge, the effective sphere size should vary with the configurations. This concept is now incorporated into the atomic ESM, as described in the next section.

\subsection{ESM with variable spheres \label{sec_vs}}
Let us look back at Eq.~(8c) in Ref.~\cite{starrett24excited}, which reads
\begin{equation}\label{eq:charge_neutrality}
    \gamma\sum_xW_x\left[\int_Vd\br n_x(r)-Z_x\right]=0\,,
\end{equation}
where $V$ is the sphere volume, which is the same for all atoms. This constraint may be understood as enforcing charge neutrality within each sphere of a fixed volume $V$. To model the effect of different sphere sizes, we introduce now the requirement that the electron densities on the surface of the spheres all have the same value, $n_e^0$.  This choice is physically motivated by the fact that, in a real plasma, electron densities on the boundary between neighboring atoms should be identical.  Enforcing this constraint, together with the assumption of spherical symmetry of the atoms, leads to the equality of the electron densities on the sphere boundaries.  Another reasonable choice would be equality of pressure.  This condition, that all electron densities at sphere surfaces be the same, is enforced variationally by replacing Eq.~\eqref{eq:charge_neutrality} with the constraint
\begin{equation}
   \gamma \sum_x W_x \left[\int d^3r (n_x(r)-n_e^0) - \Delta Z_x \right] = 0\,,\label{eq_c3}
\end{equation}
where $\Delta Z_x = Z_x - n_e^0 V_x$ and $V_x$ is the volume of the sphere associated with excited state $x$. With this change, only the minimization of the total free energy with respect to $W_x$ changes, resulting in
\begin{equation}
\begin{split}
     W_x = & \frac{ \exp\left[-(\Delta F_x - \gamma \Delta Z_x)/T  \right]} {\sum_x \exp\left[-(\Delta F_x - \gamma \Delta Z_x)/T  \right]}\,,\\
     \label{eq_wx}
\end{split}
\end{equation}
where $\Delta F_x = F_x - f^0[n_e^0] V_x$ and $f^0[n_e^0] $ is the electron gas free energy density (including ideal and exchange-correlation terms). The volumes $V_x$ are required to satisfy
\begin{equation}
   V = \sum_x W_x V_x\,, \label{eq_vol}
\end{equation}
where $V$ is the average volume per atom, which is an input of the computation.  Thus, this model requires us to calculate the electron density as a function of the volume of each configuration, and to find the set of volumes such that the electron densities on the surfaces of the spheres are the same, and that the volume constraint, Eq.~\eqref{eq_vol}, is satisfied. Note that in minimizing the ensemble free energy within the new Variable-spheres model, one finds
\begin{equation}
   \gamma = V^{xc}[n_e^0]\,,
\end{equation}
and 
\begin{equation}
    P_e = -\left. \partial F/\partial V \right|_{T,N} =  - f^0[n_e^0] + n_e^0 ( \gamma + \mu) \,,
     \label{eq_pressure}
\end{equation}
where $\mu$ is the chemical potential of the free electron gas of density $n_e^0$.  In deriving this expression, as in reference \cite{starrett24excited}, we have brushed over a basic inconsistency that is also inherent in many average atom models; the wavefunctions are normalised over all space but the energy involves integrals over the sphere volumes.  This means that the pressure calculated from this formula will not be be exactly thermodynamically consistent, though experience suggests that where Friedel oscillations\footnote{Friedel oscillations occur at high densities and low  temperatures.} are damped (which is the case for most plasmas of interest), thermodynamic consistency will hold.


In the second column of Table~\ref{tab_smallex}, we present the results of applying the Variable-spheres formalism to the simple example in Sec.~\ref{sec_nxi}. One observes that the pressure and average electron density are unchanged from the One-Sphere model, but the balance of probabilities $W_x$ of the excited states has changed. This is expected to impact the calculation of plasma opacities.

\section{Procedure for Calculating Absorption Coefficient \label{sec_calca}}
In the previous section, we have reproduced the formulation of the ESM developed in Ref.~\cite{starrett24excited}. We have also introduced an improvement to the original ESM by allowing for spheres with variable sizes representing excited states in the plasma. In this section, we discuss the application of ESM to the computation of plasma opacities.

There are at least two different approaches to calculating an absorption coefficient based on the ESM.  The first is to use the one-electron energies and orbitals from the excited state calculations to directly evaluate the excitation energies and oscillator strengths.  This option has the drawback of being difficult to apply to bound-free transitions, for, in principle, one would need to resolve the excited states in the continuum. This route is thus impractical due to the infinite number of states in the continuum. 

Another option is to calculate the independent response based on each excited state solution and correct the single-particle energy differences using the calculated excited state energies, while the bound-free threshold is calculated similarly, but removing the thermalization energy in the final state~\cite{thelen24}.  This is the approach that we take here.  It is somewhat inconsistent as the line strengths come from the one-particle matrix elements, while the excitation energies come from the difference of excited state energies, but it is a reasonable place to start.  A more consistent approach could mirror time-dependent density functional theory, but for the ESM.  Such an approach remains to be developed.

In the present approach, the frequency-dependent absorption coefficient $\alpha(\omega)$ is given by
\begin{equation}
\alpha(\omega)= \frac{ 4\pi 
\sigma(\omega) }{n(\omega) c}\,,
\end{equation}
where $c$ is the speed of light, $n(\omega)$ is the index of refraction, and $\sigma$ is the ac conductivity
\begin{equation}
\sigma(\omega) = \sum_x \sigma_x(\omega)\,,
\end{equation}
with
\begin{align}\label{eq:sigma}
\sigma_x(\omega) &= \frac{2\pi e^2}{3\omega V_x}\int d\epsilon d\epsilon'
\sum_{ll'} 
n_{x\epsilon l}(1-n_{x\epsilon'l'}) (1-e^{-\frac{\hbar\omega}{kT}})\nonumber\\
&\times \delta(\epsilon_{xl} - \epsilon'_{xl'} -\hbar \omega)
r_{x\epsilon'l'\epsilon l}^2 
\left(
l \delta_{l',l-1}+
l' \delta_{l',l+1}  
\right)\,,
\end{align}
and
\begin{equation}
r_{x\epsilon'l'\epsilon l} =(\epsilon'_{xl'} - \epsilon_{xl})
\int rdr y_{x\epsilon l}(r) y_{x\epsilon'l'}(r) \,.
\end{equation} 
Note that in Eq.~\eqref{eq:sigma}, we have the excited-state-dependent volume $V_x$, indicating the use of the Variable-spheres formulation. Note also that in this work we will use the terms absorption and opacity interchangeably, though strictly speaking, opacity includes photon scattering which is not included here.

In a practical computation, one replaces the eigenvalues with the corresponding initial or final excited state energy, if such a state is available in the list of excited states. Further, one replaces the direct and exchange energies of the bound electrons in an excited state configuration, calculated self-consistently using the LDA, with the corresponding Hartree-Fock (HF) energies~\cite{piron13}. This is done since it is well-known that the LDA produces significant errors for the energies of bound orbitals whereas HF energies are generally in good agreement with experimental results\cite{wilson95}.  It is possible that improved functionals (e.g., Becke-Johnson~\cite{RN489,RN490}) could remove the need for this \textit{ad hoc} replacement of energies. For our current purpose of demonstrating the capability of ESM, however, this simple technique suffices.  Lastly, we note that temperature-dependent exchange and correlation \cite{ksdt} effects could be tested within the context of the current model.

\section{Results \label{sec_res}}
In Sec.~\ref{sec_esm}, we have summarized the Excited States Method (ESM)~\cite{starrett24excited}, presented an extension to this method wherein the variable sizes of the atoms in the plasma are taken in to account, and described the procedure for computing plasma opacities using ESM. In this section, we show the results of our calculations for the opacities of an oxygen plasma at mass density 0.11 g/cm$^3$ and temperature 175 eV and a silicon plasma at mass density 0.045 g/cm$^3$ (areal density 80 $\mu$g/cm$^3$) and temperature of 60 eV. Unless mentioned otherwise, all computations are done with ESM$+$Variable-sphere.

\subsection{Oxygen \label{sec_oxy}}

\begin{figure}
\begin{center}
\includegraphics[scale=1]{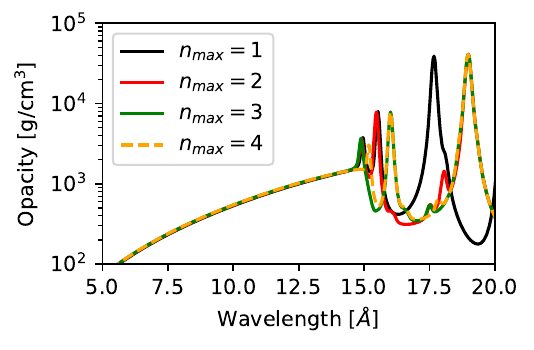}
\end{center}
\caption{Convergence of the opacity of an oxygen plasma (0.11 g/cm$^3$ and temperature is 175 eV) with respect to the maximum principal quantum number $n_{\rm max}$ of integer occupied states.}
\label{fig_converge}
\end{figure}

We begin our presentation with a discussion on the convergence of our calculations. In Fig.~\ref{fig_converge}, we show our computations for the opacity of oxygen with different values of $n_{\rm max}$, where $n_{\rm max}$ is the highest principal quantum number included in the core (one-electron orbitals with integer occupation numbers); all higher $n$ orbitals are occupied with Fermi-Dirac occupations~\eqref{eq:n_FD}. For all opacity results shown, we include all possible non-degenerate configurations which result from integer permutations of electrons in shells up to and including $n_{\rm max}$ \cite{aberg2024note}.

The strongest peak on the $n_{\rm max}=1$ curve may be associated with the $1s \to 2p$ transition, but its center is not in the correct place (not converged). This is because the final configuration $1s^02s^02p^1$ is not available in an $n_{\rm max}=1$ calculation, which only includes $1s^0$, $1s^1$ and $1s^2$. In the $n_{\rm max}=2$ calculation, the final configuration $1s^02s^02p^1$ is available, and the $1s \to 2p$ line is converged, i.e., its position shows no visible change as one increases $n_{\rm max}$ to 3 and 4. 

It is worth noting that, strictly speaking, the state $1s^02s^02p^1+$FD in the model with $n_{\rm max}=2$ does not have the exactly same energy as the state $1s^02s^0 2p^1 3s^0 3p^0 3d^0 + $FD in the $n_{\rm max}=3$ model or the state $1s^02s^0 2p^1 3s^0 3p^0 3d^0 4s^0 4p^0 4d^0 4f^0 + $FD in the $n_{\rm max}=4$ model, since the FD component in the $n_{\rm max}=2$ model has some occupation of the $n=3$ shell. This, however, is clearly a small effect, since the $1s \to 2p$ feature is well converged by $n_{\rm max}=2$. The second strongest peak on the oxygen opacity curve corresponds to a $1s\to 3p$ transition, which starts to converge for $n_{\rm max}=3$, for reasons similar to that discussed above.  The final peak corresponds to the $1s\to 4p$ transition, which starts to converge for $n_{\rm max}=4$. As indicated by the average atom model \texttt{Tartarus}~\cite{starrett19}, an oxygen plasma at the conditions under consideration only has bound states up to and including $n=4$. To include $n_{\rm max}=5$ or higher is, of course, possible in the ESM, but this would be a poor approximation as these higher-$n$ states lie in the continuum.

\begin{figure}
\begin{center}
\includegraphics[scale=1]{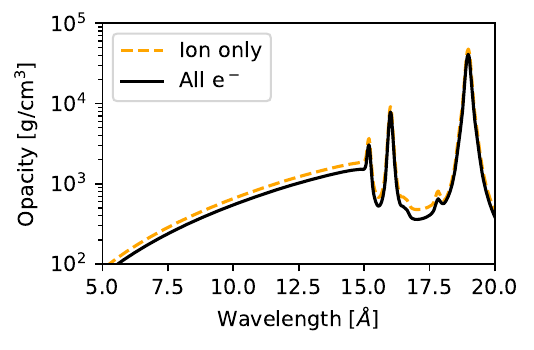}
\end{center}
\caption{The effect of including free electrons versus using bound electrons only in an ESM calculation of oxygen opacity at mass density 0.11 g/cm$^3$ and temperature 175 eV.}
\label{fig_ion}
\end{figure}
Next, let us look at the effects on the opacity due to different treatments of the free electrons.  In Fig.~\ref{fig_ion}, we show the opacity of oxygen with $n_{\rm max}=4$.  The curve labeled ``All e$^-$'' corresponds to the full model, whereas the ``Ion only'' curve results from a model that ignores the entropy and energy of the free electrons, i.e., only the bound-electron contributions to the free energies ($F_x$ and $\Delta F_x$) are included. One observes that the ``Ion only'' result is $\approx 17 \%$ higher in the bound-free region. This difference stems from a 17\% increase in the population $W_x$ of the $1s^1 2s^0 2p^0 3s^0 3p^0 3d^0 +$FD configuration, from $W_x =$ 0.284 to $W_x = 0.331$, when going from ``All e$^-$'' to ``Ion only''. The point of this discussion is to demonstrate the importance of properly treating free electrons: while the ``Ion only'' approximation is internally consistent and sound, the use of it leads to relatively poor results.

\begin{figure}
\begin{center}
\includegraphics[scale=1]{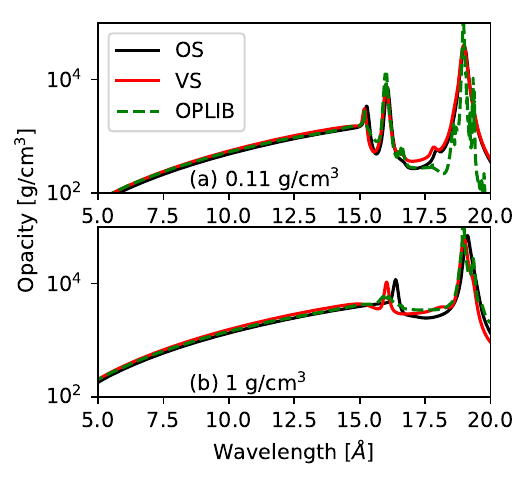}
\end{center}
\caption{Opacities of oxygen plasmas at a temperature 175 eV and densities 0.11 g/cm$^3$ and 1 g/cm$^3$. The label OS indicates ESM with a single fixed sphere size, VS indicates ESM with variable sphere sizes, and OPLIB is the Los Alamos opacity database.}
\label{fig_solar}
\end{figure}
With these technical points out of the way, we now discuss our main results. In Fig.~\ref{fig_solar} we show and compare the oxygen opacities at 0.11~g/cm$^3$ and 1~g/cm$^3$ from ESM+Variable-sphere (VS), ESM+One-sphere (OS), and the OPLIB database~\cite{colgan16}.  We see that in the bound-free region, the VS model gives a higher opacity than the OS model for both densities.  The VS result is also higher than the OPLIB database near the bound-free threshold, although the general agreement is very good.  On the other hand, for 0.11 g/cm$^3$, at wavelength $\sim$14 \AA{} the VS model is 10\% higher than the OPLIB results, whereas for density 1 g/cm$^3$, the difference between the two models at 14 \AA{} is $\sim$13\%. An increase of this magnitude in the oxygen opacity near solar interior conditions (0.11 g/cm$^3$, 175 eV) could significantly affect energy transport in stars \cite{basu2008helioseismology, serenelli2009new}. However, it should be pointed out that in obtaining these opacities, we have applied an arbitrary Voigt broadening with a width of 2.5~eV to the raw spectra. This broadening obfuscates differences between the shape of the bound-bound features when comparing OPLIB to VS and may be too crude to allow drawing any definite conclusion. A more sophisticated, consistent treatment of line shapes is the subject of another study.  

\begin{figure}
\begin{center}
\includegraphics[scale=1]{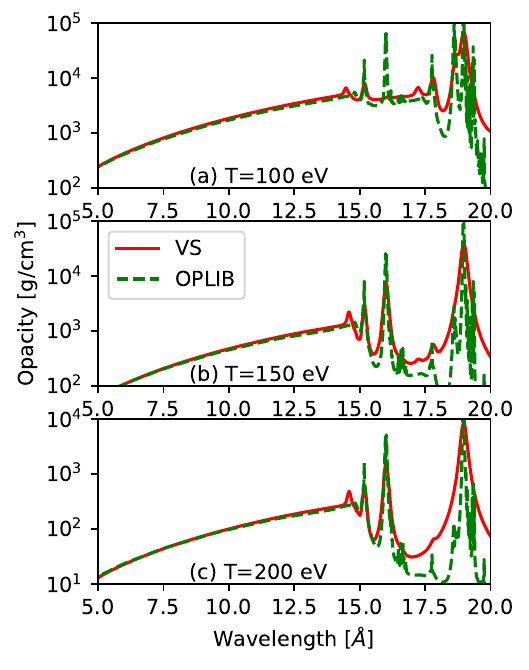}
\end{center}
\caption{Oxygen opacity along an isochore at 0.03 g/cm$^3$ for temperatures 100, 150, and 200 eV. }
\label{fig_isoc}
\end{figure}
In Fig. \ref{fig_isoc}, we show the opacities of oxygen at 0.03~g/cm$^3$ \footnote{This density is close to the density of ongoing experiments on the Z-machine at Sandia National Laboratories \cite{Mayes23}} and at three temperatures 100, 150 and 200 eV.  Again, the overall agreement between OPLIB and VS is good, but there are some differences in the bound-free region. Differences on the bound-bound lines are, again, obscured due to the arbitrary broadening in the VS results.  At wavelength 14 \AA{}, the VS opacity is 11\% (for temperatures 100 and 150 eV) and 9\% (for temperature 200 eV) higher than the OPLIB opacity.

\subsection{Silicon \label{sec_si}}
\begin{figure}
\begin{center}
\includegraphics[scale=1]{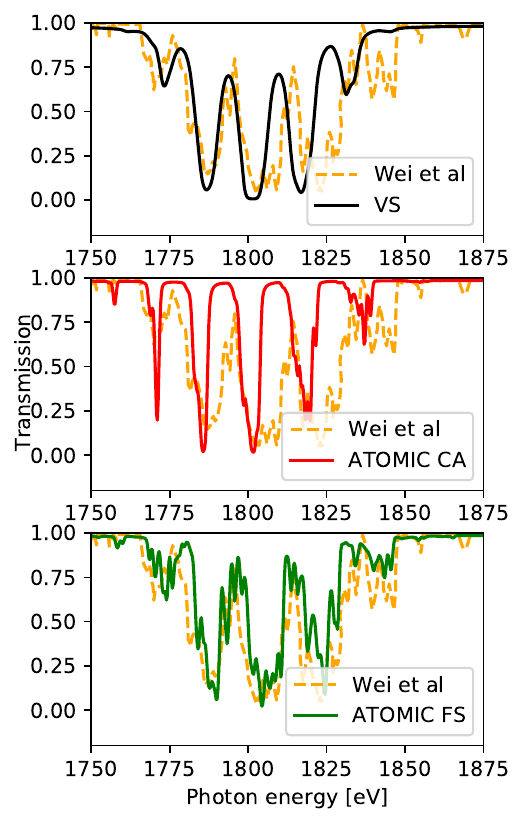}
\end{center}
\caption{Comparison between silicon opacities (0.045 g/cm$^3$ and 60 eV) from ESM with variable sphere sizes (VS) and $n_{\rm max}=2$ and from experiment by Wei \textit{et al.}~\cite{wei2008opacity}.  Also shown are calculations generated with \texttt{ATOMIC} from the Los Alamos suite of codes~\cite{colgan16, fontes2015alamos}, using configuration-average (CA) and fine-structure (FS) modes. }
\label{fig_wei}
\end{figure}
Let us now consider a silicon plasma at density 0.045~g/cm$^3$ (areal density 80 $\mu$g/cm$^3$) and temperature 60 eV. These plasma conditions are similar to the 2008 experiment by Wei \textit{et al.}, see Ref.~\cite{wei2008opacity}. The computation of silicon opacity is more challenging than that for oxygen described above. This is because for oxygen, we only need to consider the possibility of 2 bound electrons (for oxygen at 0.11g/cm$^3$ and 175 eV, the \texttt{Tartarus} average ionization is 7.5, compared to the neutral state of 8 electrons), whereas for silicon, one needs to consider ions with up to 10 bound electrons (for silicon at 0.045 g/cm$^3$ and 60 eV, the \texttt{Tartarus} average ionization is 8.2, compared to the neutral state of 14 electrons). This situation quickly leads to the proliferation of bound state configurations from which we can construct the excited states. However, since the experiment~\cite{wei2008opacity} measured the $1s\to 2p$ transition, to limit computational expense, we restrict ourselves to an $n_{\rm max}=2$ model. In Fig.~\ref{fig_wei}, we plot the plasma transmission of the silicon plasma as computed using ESM+Variable-sphere (VS) (solid black curve) and compare it with experimental result~\cite{wei2008opacity} (dashed yellow curve). Each dip in Fig.~\ref{fig_wei} corresponds to a different charge state of silicon, with the rightmost peak corresponding to $\sim$3 bound electrons, the next one to its left corresponding to $\sim$4 bound electrons, and so on. For dips on the left side of Fig.~\ref{fig_wei}, we observe reasonably good agreement between the VS calculation and experimental result. As one moves right (fewer bound electrons), the agreement deteriorates, echoing a finding made in Ref.~\cite{thelen24}.  

To investigate this, we used the \texttt{ATOMIC} plasma kinetics and spectral modeling code \cite{colgan16,fontes2015alamos} to run a configuration-resolved calculation (known as configuration-average (CA)), and a fine-structure (FS) calculation, figure \ref{fig_wei}.  The configuration-average calculation can be viewed as being at the same level of approximation as our model, in the sense that there is no angular momentum coupling, which produces more refined energy levels associated with the concepts of intermediate coupling and configuration interaction (see \cite{fontes2015alamos} for details).  However, the models are very different.  One salient difference is that \texttt{ATOMIC} does not have a self-consistent treatment of free electrons or self-consistent populations, in contrast to the present model.  Nevertheless, agreement on line positions between \texttt{ATOMIC}-CA and the present model is generally good.  Turning on the fine-structure option in \texttt{ATOMIC} (\texttt{ATOMIC}-FS) \cite{colgan16,fontes2015alamos}, the positions of the lines move to be in better agreement with the experiment.  We note that these particular fine-structure calculations arise from OPLIB calculations that employ full configuration interaction for charge states with three or less bound electrons, while the more approximate method of intermediate coupling is employed for the lower charge states, due to computational expense.  We therefore identify intermediate coupling and configuration interaction as the missing effects from the present model that are predominantly responsible for the observed differences in line positions.  

In principle, better transition energies could be found by an improved exchange and correlation functional (rather than going to configuration interaction).  This avenue remains to be explored.

\section{Conclusions\label{sec_con}}
In this paper, we introduced an improvement to the excited states method (ESM) originally developed in Ref.~\cite{starrett24excited}. This upgrade accounts for the fact that more highly charged species will have, on average, nearest neighbors farther away than lesser charged species. This effect is incorporated by allowing the atomic sphere size to depend on the ion charge, enforcing the conditions that the average volume per ion is maintained and that the electron densities on the edge of the spheres are the same across the plasma. Calculations of plasma opacities based on ESM with variable spheres have been also been presented and their predictions compared with those from other state-of-the-art opacity methods and experiment.

It was found that for oxygen plasmas at near solar interior conditions, ESM line positions (i.e., excitation energies) closely agree with those contained in the OPLIB database~\cite{colgan16}, whereas the ESM bound-free opacities are somewhat higher than the OPLIB values. We attribute this discrepancy to the ESM inclusion of free electrons, which is absent in OPLIB \cite{colgan16}. Note also that variations in the entropy of the free electrons for different excited states also affect the populations of ions. This effect has not, to our knowledge, been included anywhere else but ESM. The results of our ESM computation for a multi-bound-electron silicon plasma were also presented and compared to experimental data for the $1s\to 2p$ transition.  We found reasonably good theory-experiment agreement for lower charged ions, but the agreement becomes poorer for lines associated with higher charged ions.  An analysis of this discrepancy with the Los Alamos \texttt{ATOMIC} spectra modeling code indicated that configuration interaction corrections to the bound state energies of the ions would improve the agreement between theory and experiment.

\section*{Acknowledgments}
LANL is operated by Triad National Security, LLC, for the National Nuclear Security Administration of the U.S. Department of Energy under Contract No.~89233218NCA000001.

\bibliographystyle{apsrev4-2}
\bibliography{phys_bib}

\end{document}